\documentclass[
]{ceurart}

\usepackage{listings}
\usepackage{tabularx}
\usepackage{subcaption}

\begin{document}

\copyrightyear{2025}
\copyrightclause{Copyright for this paper by its authors.
  Use permitted under Creative Commons License Attribution 4.0
  International (CC BY 4.0).}

\conference{RAFFI-el: Workshop on Robotics and Arts for the Future of the Food Industry, September 10, 2025, Naples, Italy}

\title{How Do Technological Prototypes in the Food Industry Impact People's Perception? Insights from the MUSAE ``GROW,  COOK, CODE" Final Exhibition}

\author[1]{Francesco Semeraro}[%
orcid=0000-0002-8812-0968,
email=francesco.semeraro.research@outlook.com,
url=https://research.manchester.ac.uk/en/persons/francesco.semeraro,
]
\cormark[1]

\address[1]{Manchester Centre for Robotics and AI, The University of Manchester, Oxford Road, M13 9PL, Manchester, UK}

\author[2]{Filip Bečanović}[%
orcid=0000-0003-2952-592X,
email=filip.becanovic@gmail.com,
url=https://www.etf.bg.ac.rs/en/faculty/staff/filip-becanovic-5849
]

\author[2]{Maja Trumić}[%
orcid=0000-0002-2945-8010,
email=maja.trumic@etf.ac.rs,
url=https://www.etf.bg.ac.rs/en/faculty/staff/maja-trumic-5830
]
\address[2]{School of Electrical Engineering, University of Belgrade, Bulevar kralja Aleksandra 73, 11120, Belgrade, Serbia}

\author[2]{Kosta Jovanović}[%
orcid=0000-0002-9029-4465,
email=kostaj@etf.rs,
url=https://www.etf.bg.ac.rs/en/faculty/staff/kosta-jovanovic-4454
]

\author[1]{Angelo Cangelosi}[%
orcid=0000-0002-4709-2243,
email=angelo.cangelosi@manchester.ac.uk,
url=https://research.manchester.ac.uk/en/persons/angelo.cangelosi,
]

\cortext[1]{Corresponding author.}

\begin{abstract}
  This work reports the results of the survey carried out during the MUSAE final exhibition to assess its impact on people's perception of aspects like trust in technology, environmental challenges, eating habits and potential increase of mental and physical health while interacting with the technological prototypes exposed during the exhibition. The results show that the exhibition positively increased people's awareness regarding these aspects.
\end{abstract}

\begin{keywords}
  MUSAE \sep
  DFA method \sep
  user study validation \sep
  food industry
\end{keywords}

\maketitle

\section{Introduction}

The main aim of the MUSAE project \cite{o2025musae} is the validation of the Design Futures Art-driven method (DFA) method. The DFA method is a design methodology used to enhance collaboration between artists and small/medium enterprises (SMEs), derived as combination of Design Futures Thinking with Art Thinking \cite{jacobs2018intersections,evans2010design}. The DFA method guides technologists and artists to initially come up with future scenarios that are then used as inspirational concepts to build a technological prototype for a future need.
Within the context of MUSAE’s second residency, 11 groups of artists and SMEs were selected to develop a technological prototype, based on future scenarios related to food industry. Starting from any of the future scenarios conceived during the first residency, they had to make use of the DFA method during their prototype development. One of the objectives of the MUSAE project is for the prototypes to reach Technology Readiness Level 5, which stands for a technology basic validation in a relevant environment \cite{mankins1995technology}. 
Part of this evaluation involves testing the prototypes with potential end users. Alongside the user validations carried out specifically by each team, it is also important to understand how such a wave of technological innovation and reconsideration of our relationship with food have potential of influencing people as a whole. For this purpose, in this work, people's perception of these prototypes was assessed at a public exhibition in which the most up-to-date versions of the prototypes were exposed altogether.

\section{Setting}

\begin{figure*} 
  \centering
  \includegraphics[width=\linewidth]{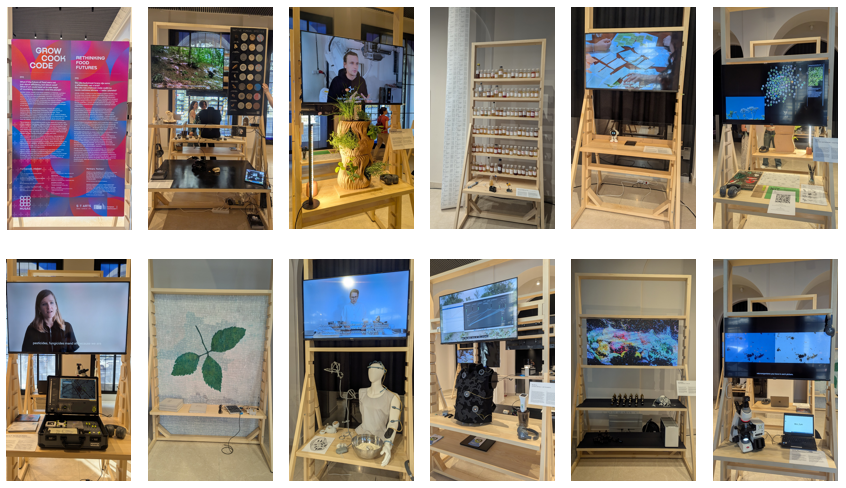}
  \caption{Pictures of the introductory panel and the 11 stands of the MUSAE final exhibition.}
  \label{exhibition}
\end{figure*}

The MUSAE ``GROW, COOK, CODE" final exhibition took place in a hall dedicated to temporary exhibitions in the ground floor of the Palace of Science (Belgrade, Serbia) from 21st to 23rd June 2025, from 11 am to 7 pm CET without interruptions. At the entrance of the exhibition, a panel describing the main features of the MUSAE project, both in English and Serbian, welcomed the visitors (see Fig. \ref{exhibition}). Then, across the entire exhibition room, 11 stands that housed the prototypes were exposed, together with a written explanation and a video of their concepts. Members of the teams who designed the prototypes were occasionally there to answer visitors’ questions. The visitors could go through the exhibition at their own pace with no specific path to follow nor time constraints. 
In this setting, the experimenter was standing next to a QR code whose scanning led to an online survey on Qualtrics, available in English and Serbian. This was inclusive of a Participant Information Sheet, consent form, questions for personal details and the survey itself. Whenever the visitors came across the experimenter, he asked them to scan the QR codes with their mobile phones so they could have the survey available to be filled. Alternatively, they had the option to request a paper version of the survey. They could then continue to experience the exhibition, also asking questions about it to the experimenter if needed. The experimenter did not check whether the visitors completed the survey thereafter. The participants did not receive any compensation for taking part in the survey.

The experiment received ethical exemption from the Department of Computer Science Ethics Committee of The University of Manchester (Ref. 2025-23115-41107). The Participant Information Sheet and the consent form were both GDPR compliant.

\section{Survey}
The survey was composed of 13 5-point Likert scale \cite{nemoto2014likert} items and one open-ended question on the exhibition. They were aimed to probe how perception of visitors was affected by their close interaction with the prototypes. Specifically, the following 4 aspects were considered:
\begin{itemize}
    \item perceived improvement of physical and mental health, imagining a projected long-term interaction with the prototypes (Health);
    \item trust in the technologies employed in the prototypes’ development (Trust);
    \item awareness of their own eating habits (Eating habits);
    \item awareness of global environmental challenges (Environment).
\end{itemize}
The full version of the survey is reported below. Some items of the survey asked participants regarding similar variables, as evidence of the consistency of their answers. The perception of the visitors regarding the aspects was computed as average of the following groups of questions:
\begin{itemize}
    \item Health: Health (P), Health (P+M) and Health (M);
    \item Trust: AI, Robotics, Wearables and Trust;
    \item Eating habits: Eating habits 1, Eating habits 2 and Eating habits 3  
    \item Environment: Environment 1, Environment 2 and Environment 3.
\end{itemize}
The open-ended question was the following: \textit{“Please write any comments about the exhibition”}. The question was purposely left as generic as possible to not bias the thoughts of the visitors. A thematic analysis \cite{harper2011qualitative} was performed on the provided answers.

\begin{table*}
\centering
\caption{The 5-point Likert scale items administered through the survey, together with their identifiers.}
  \label{tab1}
    \begin{tabularx}{\textwidth}{@{} l X @{}}
\toprule
\textbf{Question Tag} & \textbf{Question Text} \\
\midrule
Health (P) & I believe the prototypes showed in the exhibition I experienced can help improve my physical health. \\
AI & My trust in artificial intelligence (AI) has increased after the exhibition. \\
Eating habits 1 & After the exhibition, I am considering changing my eating habits. \\
Environment 1 & I have become more willing to take personal action to address environmental issues. \\
Eating habits 2 & The exhibition has made me reflect on the possibility of changing my eating habits. \\
Health (P+M) & Overall, the prototypes shown in the exhibition have the potential to enhance both physical and mental wellbeing. \\
Robotics & My trust in robotics has increased after interacting with the prototypes of the exhibition. \\
Environment 2 & My attitude towards environmental challenges has changed after the exhibition. \\
Wearables & My trust in wearable devices has increased after the exhibition. \\
Eating habits 3 & The prototypes shown at the exhibition have encouraged me to think about modifying my eating habits. \\
Health (M) & I believe the prototypes I experienced during the exhibition can help improve my mental health. \\
Trust & Overall, my trust in new technologies has increased. \\
Environment 3 & I am now more concerned about climate change than I was before. \\
\bottomrule
\end{tabularx}
\end{table*}

\section{Results and Discussions}

\begin{figure*}
    \centering
    \begin{subfigure}[b]{\textwidth}
        \centering

\includegraphics[scale=0.55]{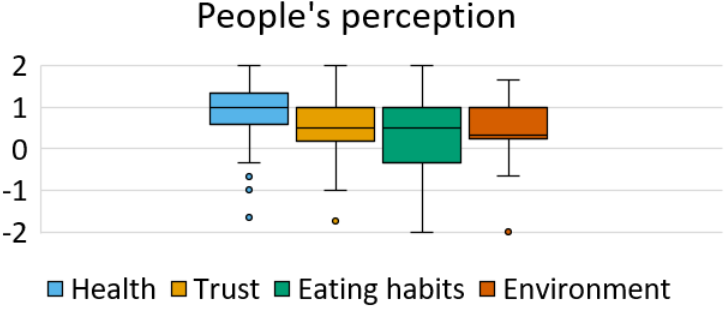}
        \caption{ }\label{aspects}
        
    \end{subfigure}
    
    \hfill
    
    \begin{subfigure}[b]{\textwidth}
        \centering

\includegraphics[scale=0.55]{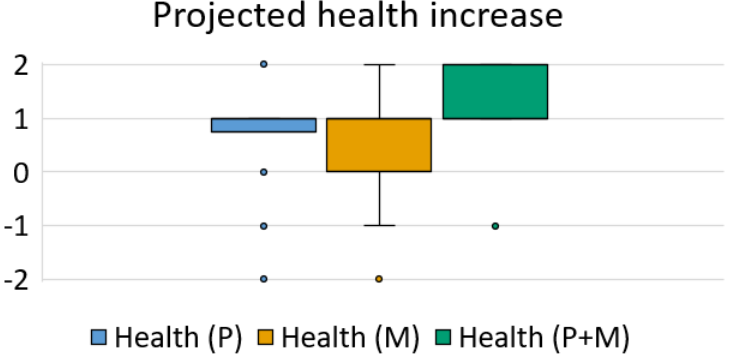}
        \caption{ }\label{health}
        
    \end{subfigure}
        
\hfill

\begin{subfigure}[b]{\textwidth}
\centering
\includegraphics[scale=0.55]{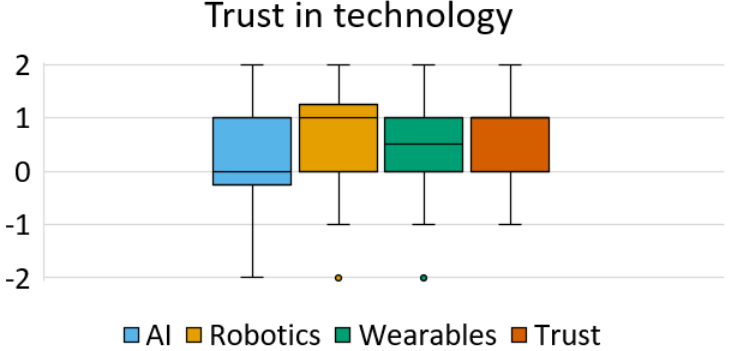}
        \caption{ }\label{trust}
    \end{subfigure}

    \caption{Distributions of scores from the survey. (a) The participants' perception on the aspects surveyed. (b) Components of the Health aspect. (c) Components of the Trust aspect.}
\end{figure*}

Throughout the exhibition, 24 fully filled surveys were collected (10 males, 14 females; age: mean 34 years, SD 9.86 years), with 21 answers to the open-ended question. Fig. \ref{aspects} shows evidence of the positive influence of the exhibition towards these thematics. Indeed, we can see that all the medians related to the aspects surveyed are greater than 0. Because of the way the statements were phrased (see Table \ref{tab1}), this results implies that the prototypes have induced people to:
\begin{itemize}
    \item think that the prototype could improve their physical and mental health in the longer run;
    \item trust more the types of technology employed in the prototypes;
    \item change their eating habits;
    \item become more concerned regarding environmental challenges, like climate change.
\end{itemize}

\begin{table*}
\centering
\caption{Thematic analysis result.}
\label{tab2}
\begin{tabular}{@{}>{\raggedright\arraybackslash}p{4.5cm} p{7.5cm} r@{}}
\toprule
\textbf{Theme} & \textbf{Description} & \textbf{\%} \\
\midrule
Positive reception        & General enjoyment, inspiration, praise & 76 \\
Technological innovation  & Use of tech for sustainability, futuristic solutions & 43 \\
Artistic and conceptual depth & Thought-provoking, interdisciplinary & 38 \\
Interaction and engagement & Artists or researchers present, inspiration to create & 24 \\
Venue and organization     & Praise for the venue and logistics & 14 \\
Critical reflections      & Critical functionality, overambition, lack of clarity & 14 \\
Inclusivity and relevance  & Projects not appealing to everyone, niche fields & 10 \\
\bottomrule
\end{tabular}
\end{table*}
This positive effect seems to have happened to a greater extent for the Health aspect than in the other ones. Indeed, for this aspect the median is the greatest one.
Looking closely at the components of the Health aspect (see Fig. \ref{health}), it is possible to appreciate that, on average, people think that the prototypes could induce both physical and mental health in the same way. Indeed, when prompted with the item related to both components at once, the overall effect is to shift the distribution towards higher values, but still preserving the median value.

Because of the way the items related to the Trust aspect were phrased (see Section 3.2), it is possible to consider different components for trust in technology, as well (see Fig. \ref{trust}). Robotics technologies played the biggest role in keeping the median value of the Trust item high. On the contrary, people were more doubtful regarding artificial intelligence. This can be explained by the fact that technologies with a higher embodiment tend to elicit higher trust in the users \cite{de2018automation}. Because robots and wearable devices have very distinctive embodiments, they are more likely to gain the trust of users than unembodied entities like artificial intelligence.

The thematic analysis carried out (see Table \ref{tab2}) shows that the impression of the exhibition was, on the whole, positive. They reported a positive impact of the exhibition, highlighting how innovative it was and how the concepts at the foundation of each prototype were deeply explored by the teams. Occasional appreciations of the venue itself were registered, too. Conversely, a minority of comments criticized the functionality of the prototypes in real-life applications, questioning their relevance to a wider public as they were considered niche topics to tackle.

\begin{acknowledgments}
  Francesco Semeraro’s, Filip Bečanović's, Maja Trumić's and Kosta Jovanović's work was supported by the Horizon MUSAE project (Ref. 101070421).

  Angelo Cangelosi’s work was partially supported by the Horizon projects MUSAE and TRAIL, the ERC Advanced project eTALK (UKRI Funded), the EPSRC CRADLE project and the US Air Force CASPER++ project.
\end{acknowledgments}

\section*{Declaration on Generative AI
}
  During the preparation of this work, the authors used ChatGPT in order to: Grammar and spelling check, Paraphrase and reword. After using this tool/service, the authors reviewed and edited the content as needed and take full responsibility for the publication’s content.

\bibliography{refs}

\end{document}